\documentclass[a4paper]{article}
\usepackage{amsmath,amssymb,booktabs}
\usepackage{calc,pstricks,graphicx,authblk}

\usepackage[parfill]{parskip}    
\usepackage{epstopdf}

\newcommand{\Figref}[1]{Figure~\ref{#1}}
\newcommand{\Cardiff}{Gravity Exploration Institute, School of Physics and Astronomy, Cardiff University, U.K.}
\newcommand{\AEI}{Max Planck Institute for Gravitational Physics, Germany}

\begin{document}

\title{Evolution of human cognition required  Einstein's gravitational waves}
\author[$\dagger$,$\ddagger$]{Bernard F.\ Schutz}
\author[$\S$]{Tsvi Piran}
\author[$\dagger$]{Patrick J.\ Sutton}
\affil[$\dagger$]{\Cardiff}
\affil[$\ddagger$]{\AEI}
\affil[$\S$]{Hebrew University of Jerusalem, Israel}

\date{}                                           

\maketitle

\begin{abstract}
We describe an unexpected anthropic fine-tuning of gravity: human cognition arose on Earth only because the laws of gravity included gravitational waves. Their link is the heat from decays of the radioactive isotopes ${}^{238}$U and ${}^{232}$Th, which were synthesized mainly in rare explosive mergers of binary neutron stars, brought about by the loss of orbital energy to gravitational radiation. This heat, released in Earth's interior, has (1) maintained plate tectonics and (2) likely helped keep Earth's iron core molten. The core's magnetic field has protected all life from annihilation by the solar wind. More surprisingly, relative brain size, a proxy for cognition, has seen two sharp increases, first for mammals and then for humans, both attributed by evolutionary biologists to adaptations to major climatic changes caused by  specific tectonic events. After the second event, the joining of North and South America, human brain size grew from chimpanzee levels to modern ones. If the laws of gravity had not included gravitational waves, humans would not be capable of studying the laws of gravity.

\end{abstract}


\section{A cognition that can understand its own evolution}
Humans appear to be the only animals that attempt to understand their own cognition and its origins. The ability to do this is often referred to as the mystery of human intelligence, but in this paper we will avoid the ambiguous term ``intelligence'' and simply address cognition generally. Humans' uniqueness and success is in any case more than just what is normally meant by individual intelligence: humans are able to solve complex problems only because they have cumulative culture and sophisticated communication. We use our intelligence to build on what previous generations have accomplished, and we then pass even more knowledge on to the next generation. This is as much a social activity as an intellectual one. Humans' social behaviors co-evolved with their brain size and analytic capabilities, so it is artificial to separate them when discussing how humans evolved these unique capabilities. The left and right brains have both been essential for human success. So it is more appropriate here to speak of the mystery of human cognition as a whole than of just our intelligence. This cognition arose through Darwinian evolution because it gave our ancestors the ability to survive in changing and challenging environments. 

We will see that some of the most important challenges for the evolution of our cognition had a surprising origin. One particular aspect of the laws of physics, Einstein's gravitational waves (Einstein 1915, 1918), contributed in an essential way to one of the most formative evolutionary challenges that humans experienced, a challenge that is believed to have driven human cognition to develop from something similar to that of chimpanzees to its present capabilities. This is by itself a remarkable association. But it is more. It goes to the heart of the philosophical issue of the ``fine-tuning'' of the laws of physics and of what is known as the Anthropic Principle (Carter 1974).

The attempt to understand ``how we got here'' over the cosmic timescale since the Big Bang has led to a long-running discussion of the multitude of ways in which the laws of physics and the values of the fundamental constants of Nature seem to be ``fine-tuned'', having relationships that are necessary for humans to have evolved. As a simple example, consider the elementary fact that electrons are much less massive than protons. This leads to atoms with electrons orbiting far from the nucleus, when compared to the size of the nucleus. When such atoms are packed in against one another, the electrons can readily interact with other nearby nuclei and with the electrons that orbit them, and this is what gives us chemistry, solid materials, and the huge variety of substances in our world. If the electron and proton had the same mass, the electrons' orbits would be tightly bound to their nucleus, chemistry would simply never have happened, and there would be no life. This then seems to be puzzling: why should the Universe have light-weight electrons? Is that an argument for intelligent creation? No, not if we understand the Weak Anthropic Principle: if, in that hypothetical Universe, the electron and proton did have the same mass, and chemistry therefore hadn't happened, then humans could not be there to puzzle over it. Only if life were possible would there be anyone to ask the question. Since we are asking it, then we should not be surprised that the Universe has enough fine-tuning to produce us.

Most of the fine-tuning that scientists have discussed is about fundamental properties that make life possible, but which don't say anything directly about the evolution of a cognition that is sophisticated enough to ask the questions about fine-tuning. In this paper, we discuss a surprising type of fine-tuning that is directly related to the evolution of human cognition: that gravity must allow gravitational waves. We shall provide evidence here to support the following chain of deductions that leads to that conclusion. 
\begin{enumerate}
\item Gravitational waves carry energy away from binary star systems and thereby bring the stars closer together. One consequence is that the there are occasional (explosive) mergers of neutron stars (themselves the collapsed cores of dead massive stars), producing short-lived optical events we call kilonovas (Metzger 2020). (The word macronova is also used in the literature for these events.) Without gravitational waves, there would be no kilonovas.
\item Kilonova explosions have been the principal source of the elements in the periodic table above iron (Lattimer and Schramm 1974,  Eichler et al 1989, Hotokezaka et al 2018). These include uranium and thorium, whose radioactive isotopes ${}^{238}$U  and ${}^{232}$Th are distributed throughout Earth's interior. Without kilonovas, the abundance of these isotopes would be much smaller, or even zero.
\item The heat being generated inside Earth by decays of these very long-lived isotopes contributes to keeping Earth's core molten and therefore in sustaining Earth's magnetic field. Without this field, Earth would look like Mars: the solar wind would have blown away the atmosphere and oceans.
\item Additionally, this heat is  believed by geophysicists to be essential in maintaining plate tectonic activity today. Without this heat, the continents would likely have frozen in place a billion or more years ago.
\item Plate tectonics has been a major driver of evolution, and in particular of cognitive evolution. The merger of North and South America to produce the Isthmus of Panama 3 million years ago created climate changes in Africa that sparked the enlargement of the human brain from chimpanzee-size to its present size. If continents had frozen in place before this, all African primates would probably have remained small-brained. Or primates might not even have evolved, because the earlier tectonic event of the splitting of Australia from Antarctica about 40 million years ago  seems to have produced a rapid increase in mammalian brain size generally. Without that event, mammals might never have reached the intelligence even of primates.
\end{enumerate}
The fact that Earth has creatures who can understand Einstein's general relativity depends on the very existence of the gravitational waves of general relativity. Gravitational waves were not necessary for the evolution of life, nor possibly even for the evolution of mammalian life, although a case has been made for the possibility that kilonova-produced iodine was indeed necessary for life on Earth, at least as we know it (Ellis et al 2024). But we shall show that gravitational waves were part of the causal chain for the final step: making humans bright enough to start understanding their own cognition. 
 
\section{Gravity the outlier}
Let's begin with the unusual nature of gravity. Gravity is an outlier. Unlike the  electromagnetic interaction, in which opposite charges attract but like charges repel, gravity normally does only one thing: it attracts everything. The other two fundamental forces of physics (called the strong and weak interactions) are also normally attractive, but they act only between specific kinds of particles, and only over very short ranges that make them important in the atomic nucleus but render them negligible even over a distance as small as the size of an ordinary atom. So over macroscopic distances, only gravity and electromagnetism are important, both getting weaker with increasing distance as $1/r^2$. 

Moreover,  only gravity attracts everything. All other particles are selective in what they interact with. Protons, for example, interact electromagnetically with other charged particles but not with neutral ones. But, because of the equivalence of mass with energy, any form of energy has its own gravitational attraction. Every particle and every field has energy, so gravity is the only universal interaction.

Inherently, gravity is very much weaker than the other forces (far weaker even than the ``weak'' interaction). But its universal attraction allows it to become competitive with the others if there is enough matter in a small enough region. It turns out that, if an object is made of ``ordinary'' material (like rocks or solid iron) and if its mass is larger than about $3\times 10^{21}$~kg (about 5\% of the mass of our Moon), then gravity will be more important than the other forces in shaping the object's form (Schutz 2003). This is the reason that planets and large moons are smooth and round, while smaller objects like asteroids, comets, and even small moons such as Mars' Phobos are irregularly shaped.

Typical objects in our day-to-day experience are so small that their mutual gravity doesn't matter at all. The only gravity we normally have to deal with in everyday life is that of Earth itself. However, where gravity is dominant, able to overwhelm electromagnetic forces, then its relentless pull knows no way of stopping. Gravity, by itself, can achieve no \textit{equilibrium}. Equilibrium requires a balance between attraction and repulsion. It is this one-way feature of gravity that has shaped our Universe, leading to the formation of planets, stars, galaxies, and ultimately black holes. 

Normal chemical matter, by contrast, is ruled by the electromagnetic interaction, whose combination of attraction and repulsion allows it to achieve equilibrium in countless different forms. The attraction of opposite charges is balanced by the repulsion of like charges, and since they come in equal numbers, they can (with the help of the exclusion principle of quantum mechanics)  reach stable arrangements for a huge variety of solid materials. In empty space, countless rocky asteroids and icy comets, smaller than our Moon, have reached equilibrium with very irregular shapes. Unless they are unlucky enough to encounter the Sun or a planet or another rock, they will just retain their form essentially forever. 

The atomic nucleus, too, is in equilibrium if the number of protons and neutrons is right. The protons, all having the same electric charge, repel each other very strongly at these small distances. But the strong interaction acts attractively among all the protons and neutrons together. If there are enough neutrons, the strong interaction will hold the nucleus together, just balancing the electromagnetic repulsion. If you push these particles too close together, the electric repulsion will get stronger and will push the nuclear structure back to its equilibrium configuration. This is how all the stable nuclei in the periodic table maintain their equilibrium. 

Since the nuclear force is short range (attracting basically only the nearest neighbor protons and neutrons) while the electromagnetic repulsion is long-range (every proton in the nucleus pushing itself away from every other one), this balance only works for small numbers of particles. The more protons there are (in other words, the higher we go in the periodic table), the higher the neutron fraction has to be in order to achieve balance. Apart from the smallest nuclei, there are no stable ones whose neutron number is much smaller than its proton number. (Stability is actually more complicated than this, since the particles have an exclusion principle with other particles of their own type. But we don't need this level of detail for our purposes here.) 

Unstable nuclei might split, some of them emitting a ${}^4$He nucleus consisting of two protons and two neutrons, a process called alpha-decay. Others might fragment into two pieces more similar in size, a process we call fission. And in yet others, where the number of neutrons is a bit larger than  actually needed for counteracting the repulsion of the protons, one neutron might disintegrate, leaving a proton behind in the nucleus and expelling an electron and a neutrino. This is called beta decay.  The result of all these competing forces and processes is that, beyond uranium in the periodic table, there are simply no stable nuclei. We'll find that we will need to pay attention to the stability and instability of nuclei in our discussion below, when we look at how uranium and thorium are created in the first place, in collisions between neutron stars that lead to kilonova explosions.

On its own, a neutron is actually not itself in equilibrium. It is an unstable particle, and this is why beta decay is available to slightly unstable nuclei. The instability of the neutron arises because the weak interaction \textendash\ which is weaker than the strong one but still very much stronger than gravity \textendash\ is repulsive at this scale. It prefers to split the neutron apart, creating a trio of free particles \textendash\ one proton, one electron, and one antineutrino \textendash\ with a big enough kick that the proton and electron escape from each other by shooting off in different directions, despite their natural electromagnetic attraction for one another. The only reason that a neutron doesn't always do this when it is being used to glue nuclei together is that the proton it would create would still be stuck in the nucleus, bound by the strong interaction, and its mutual electric repulsion to other protons would require more energy to create it than is available from the decay of the neutron. Beta decay happens when there are still enough remaining neutrons that the proton left behind is bound despite the repulsion of the other protons.

The large variety of ways in which all three interactions (excluding gravity) work together, as hinted above, ensures that equilibrium of matter is possible in a wide variety of forms and circumstances. Our material world is made of such stable equilibrium structures. Stability implies that they are in a state of minimum energy: at least for small disturbances, to disturb the equilibrium requires energy input. A nudge to an equilibrium system, adding energy to it, typically leads to small extra motions, increasing the system's temperature. We say that such equilibria have positive specific heat: the temperature goes up when one adds energy. 

That may sound like it should always happen, but gravity is just the opposite! If you add energy to a satellite in orbit around Earth, say by firing its rockets to give it a forward push, it will move further from the planet and slow down! If you add energy to the atmosphere of a star, it will expand so that its outer regions actually get cooler. Systems controlled by their internal gravity have {\em negative} specific heat. 

We call anything big enough for gravity to dominate its structure \textit{self-gravitating}, and such bodies exist in an uneasy balance between the equilibrium-seeking of the three other interactions and the persistence of gravity in destroying the equilibrium. If a system's energy keeps decreasing (e.g.\ when a star radiates light), the equilibrium between gravity and other forces may keep readjusting and persist for a long time, but the continued contraction of the system and concomitant strengthening of its self-gravity can ultimately lead to catastrophe.

Of course, reassuring as it is to be surrounded by structures in equilibrium, our own biological existence is {\em not} an equilibrium state. At an individual organism's level, life is a process of development and change and renewal, ultimately followed by disintegration. At the species level, Darwinian evolution can sometimes come to an equilibrium when averaged over the fates of many individuals, provided its ecological niche is itself stable over time. But if conditions change  enough, then such a quasi-equilibrium can give way to cut-throat competition for survival. We will come back to this below when we consider the evolution of higher cognition.

Now, if life and its evolution require disequilibrium, then this can only have happened through the influence of gravity. To understand the importance of gravity, let's indulge in a fantasy. Imagine a Big Bang like ours, only without gravity: everything shoots out from the initial singularity, expanding rapidly, cooling, forming the elementary particles. The expanding cloud of gas would, as in our real world, consist mainly of what we call dark matter. The normal (non-dark) matter in this still very hot cloud would  \textendash\ via the nuclear interactions \textendash\ form helium nuclei and traces of lithium (the third element in the periodic table). No bigger nuclei would have enough time to form as the gas expanded and cooled. 

So far everything in this fantasy Universe is the same as in our real one. But, in the absence of gravity, after these light nuclei had formed, then nothing more would happen. No cosmic structures, no planets, no DNA. Just a cloud of simple elements expanding, getting colder and colder. The dark matter would have had small density irregularities, but these would have made no difference, since dark matter does not interact with normal matter in any way in this gravity-free Big Bang.

So the immense diversity and complexity of our Universe owes itself to gravity. Not to the overall smooth gravity of cosmology, which determines the slowing down or speeding up of the expansion of the Universe on the largest scales. Where gravity started to be important for structure was in the small random density fluctuations in the dark matter that emerged from the Big Bang, where regions of larger density had stronger gravity that was sometimes strong enough to reverse the overall cosmic expansion in that region.  Since there is ten times as much dark matter as ordinary matter, the extra gravity of the dark matter managed to put the brakes on the expansion of everything in many local regions, slowing these regions down until they re-collapsed in on themselves. Given gravity's inability to find an equilibrium all by itself, and the way it gets stronger and stronger as matter gets more and more dense, the formation of complex hierarchical structures was then inevitable. 

\begin{figure}
\includegraphics[width=3in]{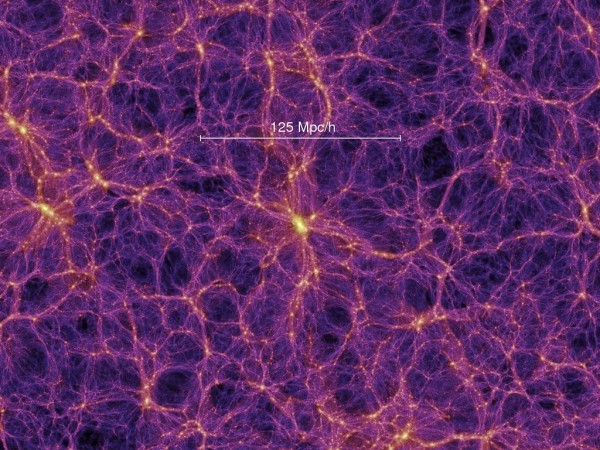}
\caption{The Cosmic Web: a supercomputer simulation of the formation of structure in the early expanding Universe. The bright spots are groups of galaxies, not stars. Credit: Volker Springel and the Virgo consortium.}
\label{fig:cosmicweb}
\end{figure}

In a process that astrophysicists have now been able to simulate on supercomputers, filaments of gas began to get denser and denser, woven together into what we call the Cosmic Web (\Figref{fig:cosmicweb}). Within these filaments,  giant clouds formed, and out of these clouds condensed the galaxies that dominate our photos of the very distant Universe. And within these galaxies, the still-collapsing gas got dense enough and hot enough for stars to form and eventually to halt the collapse by igniting nuclear reactions in their cores. The road to the evolution of life had opened up. 

\section{Gravity the midwife}

The story so far needs only what we call Newtonian gravity, the form of gravity that adequately explains our Solar System's planetary orbits. As we have described earlier, Newton's law of universal gravitation is a rule that states that every body in the universe attracts every other body with a force proportional to the inverse square of the distance between them. But Newton had no proposal for the mechanism by which this force acts; physicists refer to it as ``spooky action at a distance'', and it was criticized from the start for that reason. 

\begin{figure}
\includegraphics[width=2in]{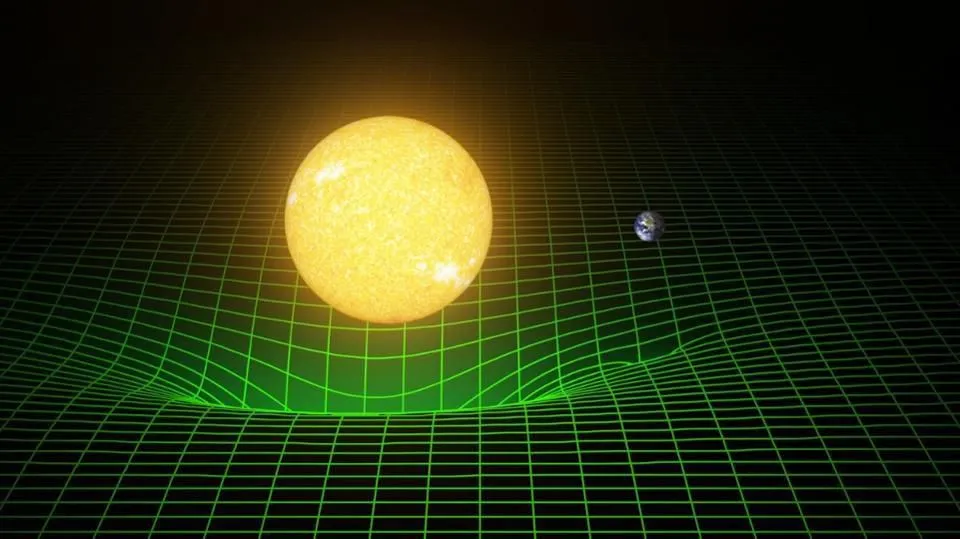}  
\includegraphics[width=2in]{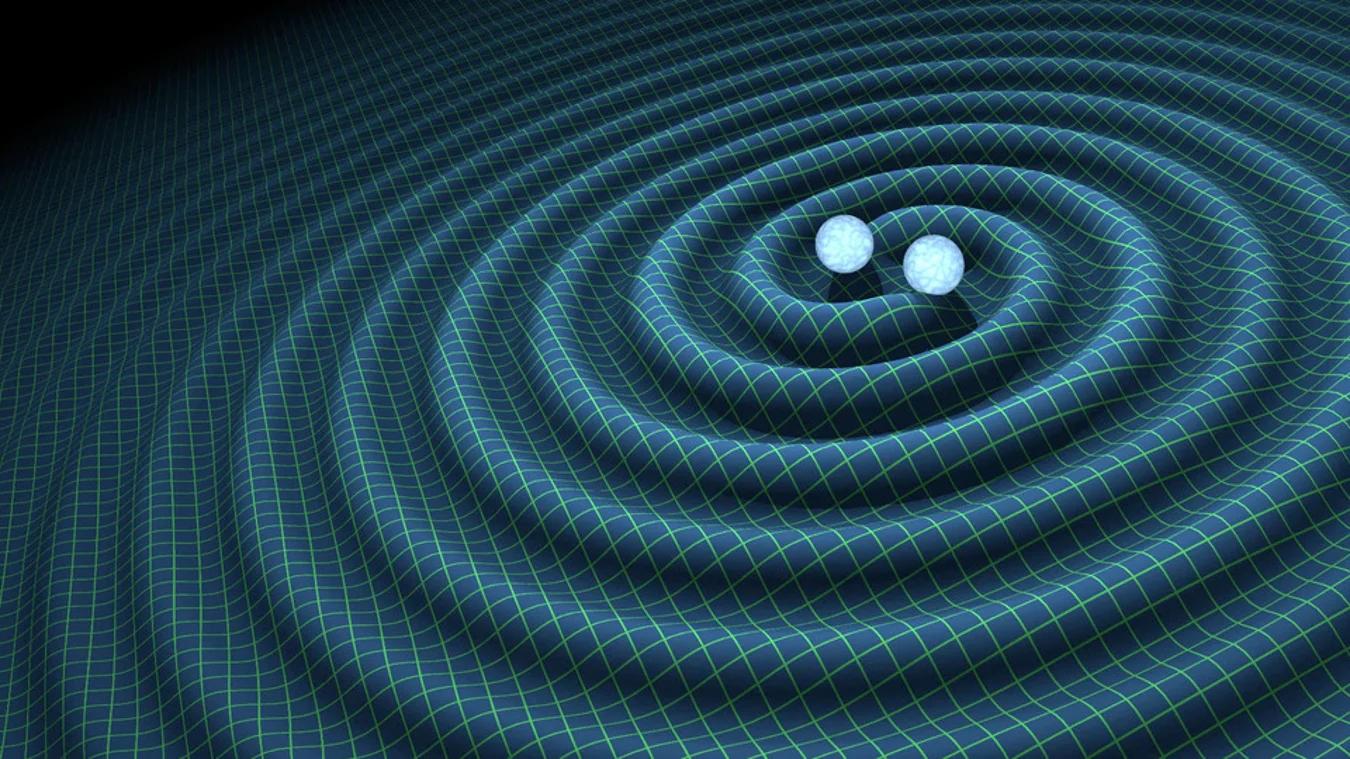} 
\caption{Left panel: The Sun changes the local geometry in such a way that 
Earth is trapped in a nearly circular orbit around it. To get away, the planet would need a higher velocity, called the escape velocity. Credit: T. Pyle/Caltech/MIT/LIGO Lab.  Right panel: Two objects in orbit around one another disturb the local geometry in a way that sends out ripples of geometry at the speed of light. These gravitational waves disturb the local gravity wherever they pass, which is how our detectors sense them. They also carry energy away, causing the objects to spiral gradually together.  Credit: LIGO Scientific Collaboration.}
\label{fig:GRGW}
\end{figure}

The most revolutionary idea in Einstein's theory of gravity (known as the general theory of relativity) was to propose that gravity can be viewed not as a force that acts mysteriously across empty space, but rather as an inherent property of space (or more correctly, of space and time). Specifically, in Einstein's theory a massive object such as the Sun warps the spacetime around it, as illustrated in the left panel of \Figref{fig:GRGW}. This warping causes other objects moving through that spacetime to follow trajectories that we perceive as curved, such as the Earth orbiting the Sun. A smaller body such as  Earth generates its own smaller warping of spacetime, which is responsible for the gravity that keeps us on the ground. In the most extreme case, black holes, the curvature becomes so strong that not even light itself can escape from the surface of the object.

Spacetime curvature also opens the possibility of waves in spacetime. The most important example for us is a close binary of two compact objects such as black holes or neutron stars, as illustrated in the right panel of \Figref{fig:GRGW}. As the two objects orbit each other, the constantly changing curvature of spacetime they produce causes waves in spacetime that radiate out from the binary at the speed of light. These gravitational waves carry energy away from the binary, causing the two objects to  approach  each other, slowly at first, but getting ever faster as they get closer and closer. 

Because these changes from Newton's gravity to Einstein's are part of relativity, they become more important when speeds get closer to the speed of light. That is why Newtonian gravity works fine in our Solar System. We need to add the complications of Einstein's gravity only when gravity gets strong enough to raise the typical speeds to a good fraction of the speed of light. So we need Einstein in order to understand the Universe  just when it was beginning, but by the time dark matter started forming structure, the temperature of the expanding gas was low enough that the random speeds of particles were non-relativistic. Therefore, what we have learned from the discussion in the previous section is that the Newtonian aspect of gravity was necessary for life to evolve.

Necessary, but clearly not sufficient: the process of forming life was extremely complex and involved all the forces of Nature, not just gravity. Many other ingredients went into the mix that is our Universe, some of them also look like they were finely tuned in order to allow a Universe with life. But our point here is to recognize that gravity was certainly necessary. Gravity in its Newtonian form has partnered with the strong and electromagnetic interactions to enable almost every stage of the path that evolution took, as far as we can discern it. The Cosmic Web formed large and small condensations of gas, and the small ones usually got trapped by the gravity of the large ones. While stars began to form in all these blobs of gas, the gas blobs themselves were also busily merging together. Most galaxies are formed by mergers of small ones falling into big ones, and sometimes by two big ones merging together.

Let us focus our attention on our own galaxy, the Milky Way. Observations by the recent European Space Agency mission called Gaia have begun to trace its merger history (Gaia Collaboration 2018). Over time, the Milky Way has cleared out its near neighborhood, so that only a few small assemblages of stars remain to be incorporated. The Small and Large Magellanic Clouds are chief among them, and are on a trajectory to merge into the Milky Way in, say, a hundred million years. A mighty collision with the Andromeda Galaxy, M31, is also on the cards, but is many billions of years away.

The first Milky Way stars that condensed out of the gas of our galaxy as it got denser and denser were not suitable hosts for life, because they consisted just of the primeval elements, hydrogen and helium. No rocky planets or moons could have formed around them. The gas cloud that formed each first-generation star contracted under the pull of its own gravity, got hotter (because of its negative specific heat), and eventually ignited nuclear reactions deep down in its center. The stars became meta-stable, the energy released by the reactions providing the pressure that stopped further collapse. These reactions mainly created more helium by using the hydrogen as fuel. A core of helium \textquotedblleft ash\textquotedblright\ built up inside, and as more ash was added to it, it got denser and hotter, until (helped by gravity's inward pull) the central temperature got hot enough to begin nuclear reactions between the helium nuclei too, allowing some of them to fuse into carbon and even heavier elements. For the first time, elements heavier than lithium began to appear in the Milky Way. 

These first stars were more massive than typical stars today, which means they evolved much more quickly. Many of them eventually ran out of fuel and just collapsed in on themselves and formed  black holes. But at least some of them exploded to become the first generation of supernovas, blowing off at least some of their newly manufactured elements. Other clouds of primeval gas in our galaxy, which had not yet formed stars, were essentially polluted by this stuff. For the first time, chemistry became possible in the Milky Way. Because we are made of these elements, astronomers don't call this pollution, they call it ``enrichment''.

The next generation of stars formed from the newly enriched gas, some of them possibly triggered to collapse by the impacts of the first supernova explosions. These did the same sorts of things as the earlier generation, but because they had the extra elements they were typically less massive (hence longer-lived). They now had two ways of expelling some of the elements they had cooked up in their interiors: the more massive of them exploded as supernovas, and many of the less massive stars expanded their outer regions as ``planetary nebulas'' or ``red giants'', eventually blowing off their exteriors and settling down into long-lived white dwarf stars. The material blown away in this way contains elements synthesized during the lifetime of the star. Supernovas, on the other hand,  blow out elements that were created even deeper inside the stars, and then synthesize even heavier elements through the nuclear reactions that happen in the dense gas during the initial phase of the explosion. All of this synthesis happens via fusion: helium nuclei  encounter  existing nuclei and merge with them to make heavier ones.  The most common basic elements were synthesized in this way: carbon, oxygen, silicon, iron, etc. The stellar winds and supernova explosions pushed them out through the Galaxy, increasingly enriching the remaining gas in the Milky Way. If rocky planets had not been possible for this second generation of stars, then they were certainly possible for the third. 

The supernovas we have been describing are actually called Type II supernovas. There is another important type, Type Ia. These are explosions of white dwarfs, the kind of star that our Sun will eventually turn into. With Type Ia supernovas we get our first glimpse of the influence of gravitational waves on nucleosynthesis. If a white dwarf forms in a binary system with another star, and if their orbital separation is close enough, then over billions of years the regular changes that their orbital motions make in the local gravitational field send out Einstein's gravitational waves, which carry energy away. The stars get closer together and eventually merge, and that merger produces the Type Ia supernova explosion. The elements that are synthesized are similar to those in Type II supernovas, and in fact they may have been the source of most of Earth's iron. They therefore contribute to the variety of chemistry on our planet, but they do not do the special things that, as we will see below, the elements produced by analogous mergers of neutron stars produce. 

At first, scientists had assumed that all the elements of the periodic table were formed this way, by fusion inside stars or in supernovas. But computer simulations have shown that this is not possible. For one thing, ${}^{56}$Fe is the most stable element, so to add a helium nucleus to it will require considerable input energy, which is not readily available inside stars. It is available in supernova explosions, but in them the process of combining elements to make heavier ones stops when the expanding gas of the supernova gets too thin, after which the existing nuclei  don't have much likelihood of colliding with others. Fusion like this  synthesizes elements in the periodic table up to iron, but not much beyond. Newtonian gravity, therefore, with a bit of assistance from gravitational waves, sufficed to midwife the lower half of the periodic table. Had there been no gravitational waves, the relative abundances of various elements would have been different, but life would surely still have been possible.

In the next section we will explain that the heavier elements (including gold, silver, iodine, uranium, etc) could not have formed with any process regulated just by Newtonian gravity, that gravitational radiation was essential to the formation of the heavy elements. But in order to understand the implications of that, we need  first to continue with the story of how our Solar System provided the nursery for the evolution of life. 

Our Sun is a late-generation star, whose progenitor gas cloud was richly seeded with all the elements heavier than lithium. The inner planets of our Solar System were hot enough to have boiled off almost all the hydrogen and helium that still bulk up Jupiter and the other outer planets. By chance, Earth found itself in the habitable zone, at a distance from the Sun where water can remain liquid; and it found itself with plenty of surface water. Somehow, whether through early electrical storms, or from cometary seeds, or from the Mars-size planet that collided with Earth early on and threw off the stuff that formed our Moon (Young et al 2016), Earth's oceans got a good dose of amino acids. Life started \textendash\ that much we know. How \textendash\ we don't know. 

Once life started, however, Darwinian evolution took over, and complex multi-cellular organisms eventually appeared. During this whole amazing story, Newtonian gravity did more than just hold the oceans onto the planet and keep our planet at just the right distance from the Sun. It also held the Sun together as a nuclear fusion furnace, providing a steady and gradually brightening flux of energy to keep the oceans liquid, and later to power photosynthesis. 

At every stage of the story so far, Newton's gravity has been present, acting as a kind of midwife to the birth of life.  Now it is time to turn to the issue of why this is not actually enough to ensure that life prospered on Earth and did not later die out, and enough to explain why  higher cognition, using bigger and more energy-expensive brains, was needed by some animals in order to pass the test of natural selection. It is time to start explaining the unique and unexpected role of Einstein's general relativity in the story of life on Earth. 

\section{The gravity that enabled production of the heaviest elements}

Astrophysicists have come to realize only fairly recently that black holes and gravitational waves are not just Einsteinian exotica, that  in fact they play an important role in many aspects of the development of the Universe that we have outlined above. What is important for evolution is that, without gravitational waves, the elements in the heavier half of the periodic table would never have been created in anything like their observed abundances. After explaining why, we will then argue that, without the particular elements uranium and thorium, it seems unlikely that cognition as advanced as that of humans  would have evolved, and it is very possible that all life would have died out by now. We begin by looking at what links black holes, neutron stars, and gravitational waves with the synthesis of the heaviest elements.

Our outline earlier of how stars form and die in supernova explosions ignored what happens to the stars after they explode. We have saved that part of the story for here because it can't be understood without Einstein's gravity. Typically, the star does not completely disintegrate in the explosion. That is because the source of the energy of the explosion is gravitational, not nuclear. Before the explosion, the star was in equilibrium between gravity and the nuclear processes: gravity pulled inward and the energy released by nuclear fusion reactions was carried by photons that pushed outwards. The equilibrium failed when the nuclear fuel started running out. The result was that the innermost core of the star, which was the only region hot enough for the nuclear reactions to take place, had difficulty supporting itself against the weight of the rest of the star. So it started to contract. To make a (very) long story short, at some point the core could no longer generate enough energy to support the rest of the star, and it collapsed inwards, pretty much in free fall. 

Although the stars that make supernovas are typically very massive, more than ten time our Sun's mass, the core when it collapses is only a bit more massive than our Sun. It turns out, quite remarkably, that when the free-falling collapsing core reaches nuclear density, it can again  develop enough pressure to stop its collapse. This is an unheard-of density: the whole solar mass of material turns into one large nucleus. Our discussion earlier of nuclear stability would suggest that this would be wildly unstable, but in this case there is an added factor: gravity. The size of this core is only 10--15~km in radius, and gravity is highly relativistic. The escape speed from such a body would be close to the speed of light. The collapse has happened at close to the speed of light, leaving the outer part of the star behind. The high escape speed means that each nuclear particle is tightly bound by gravity, because to get away at close to the speed of light it would have to have kinetic energy that is a good fraction of its entire rest-mass energy. Because that much energy is not available, the core is now a stable body. We call it a neutron star, because almost all the electrons in the collapsing core have been pushed into merger with the protons, forming neutrons that remain stable in this strong gravitational field.

Now, the collapse of this much mass into a stable configuration that is so much smaller releases a huge amount of energy, something like 10\% of the rest mass-energy of the material in the core. A tiny amount of this energy is carried away from the core by the elusive particles we call neutrinos, that are emitted when electrons and protons combine to make the neutrons. These particles have a tiny rest-mass, and the amount of energy they get when emitted by this reaction is far more than their rest-mass energy, which means that they {\em will} escape the core's gravity. But most of the energy is just the gravitational energy released by the collapse of the core, and that becomes thermal energy in the hot cloud of gas just outside the neutron star. This gas consists mostly of neutrinos and their anti-particle companions, antineutrinos, which are created in pairs when very energetic photons (hyper-energetic gamma-rays) collide with one another. This turbulent soup of photons and neutrinos has a huge pressure, which pushes outwards,  encountering the remainder of the star,  which is starting to collapse inwards as well. We call this remainder the ``envelope'' of the star. At the inner edge of the envelope, a shock develops. What happens next depends on how massive the original star was. 

In most of the stars that started out with enough mass to collapse, the envelope (which contains 90\% or more of the star's original mass) is simply blown away by all the energy released in the collapse. That is the classic supernova explosion, called a Type II supernova. The inner part of the envelope had already got very dense before encountering the neutrino gas pushing outwards, so the neutrinos and the high density and temperature there induce nuclear reactions that fuse the existing elements into even heavier ones, up to iron. All of this is blown out through the Galaxy, mixing into the clouds of hydrogen that have not yet formed stars. What is left behind is a neutron star, initially very hot, with a mass a bit larger than that of the Sun, crammed into a sphere that is only the size of a big city. 

It should not be surprising that some stars are too heavy for this scenario: the inner envelope can be too massive  to be fully blown away, so some of it falls back onto the neutron star core. If the core then grows much bigger than about twice the Sun's mass,  the nuclear forces won't be able to support the core against the increased gravity, and the core will collapse. There is no further stopping point after that: the core collapses to a black hole. The hole really is a hole: there is no longer any outward pressure on the envelope from neutrinos and the hot $\gamma$-\ and X-rays that had been given off by the core, so now matter just falls in. The outer part of the envelope had still been blown away by the neutrinos, creating a supernova explosion, but the central remnant is now a black hole with a mass much larger than that of the Sun. 

Fascinating as black holes are, they do not play a central role in our story. They are dead, unable to participate further in processes like nuclear physics. But neutron stars are just enormous nuclei. These remnants of the smaller supernovas turn out to be ingredients in the recipe for the synthesis of the heavier elements, and the evolution of human cognition.

\begin{figure}
\includegraphics[height=3in]{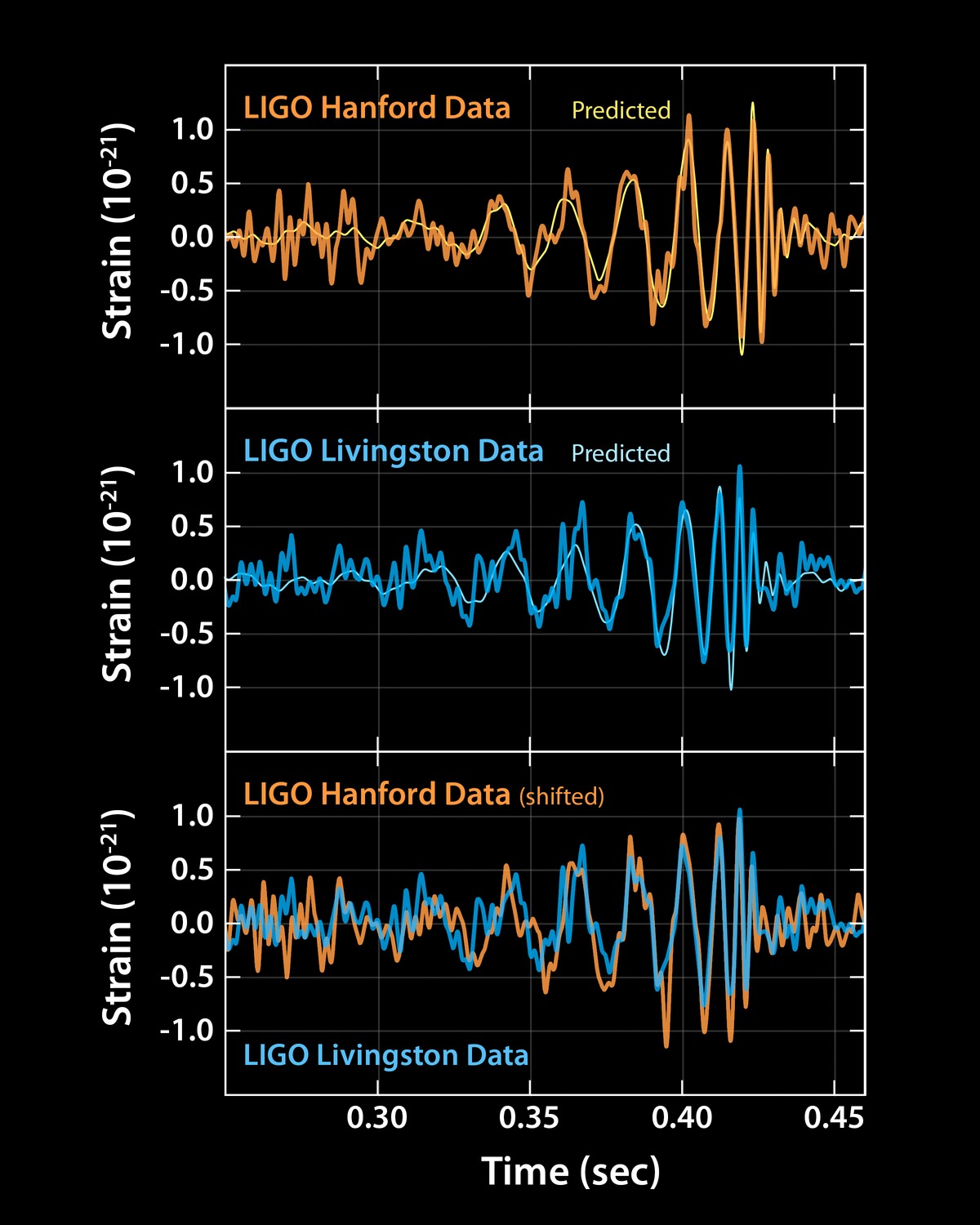}  
\includegraphics[height=3in]{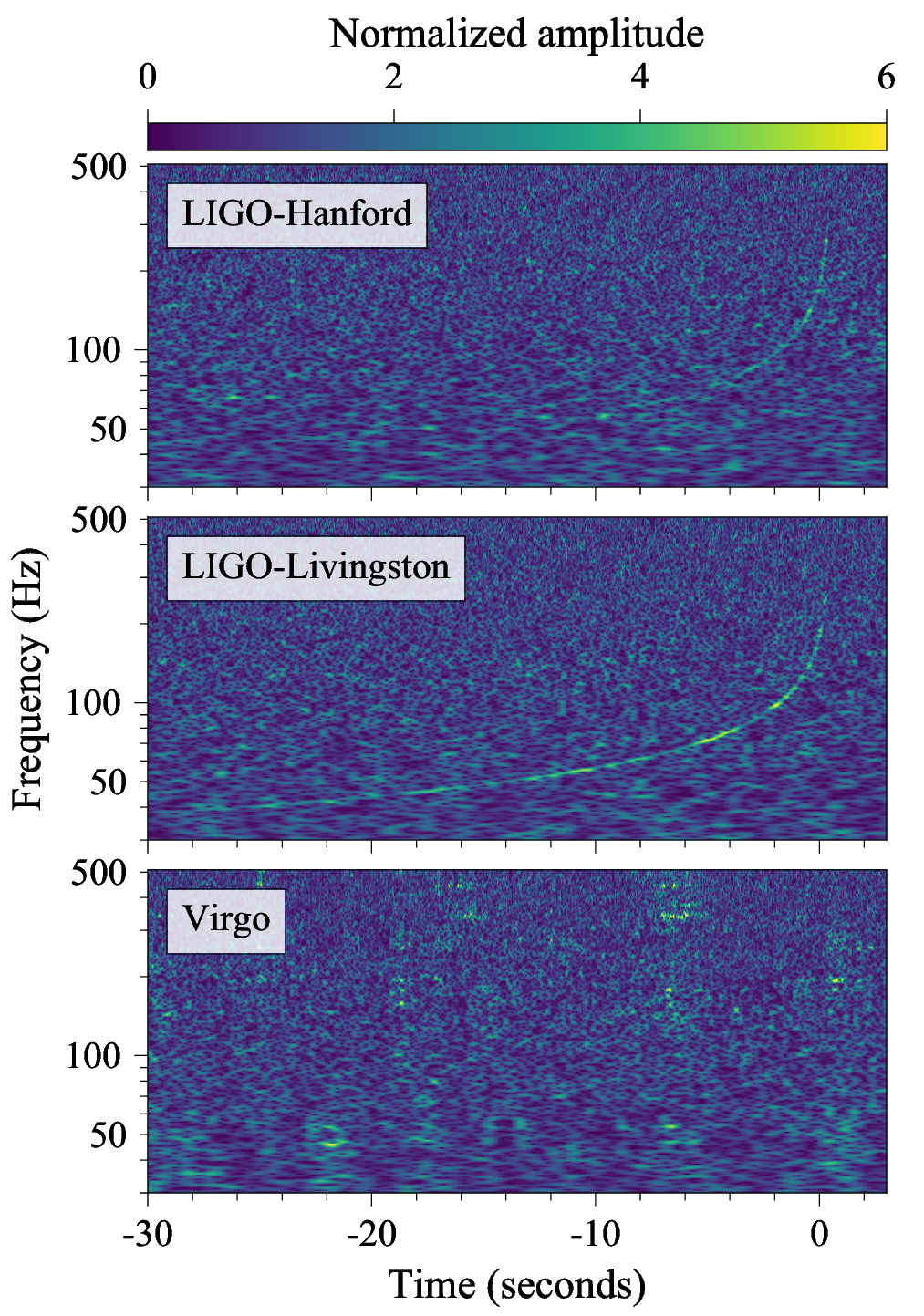} 
\caption{Left panel: The signal received by the two LIGO detectors on 14 September 2015 (top two panels), and the two signals superimposed with a slight shift in time because the signal arrived at one detector before the other.  Right panel: The signal in LIGO and Virgo of the binary neutron star merger GW170817. Since it was too weak to see the raw waveform clearly in a time-series like the one on the left, this panel shows the loudness of the signal as it would be heard by the ear, plotted against time. The pitch is plotted vertically, and loudness is indicated by the colour.  It was weak in Virgo because its location on the sky happened to be in a direction to which Virgo had very low sensitivity. Notice the very big difference in the time-scale of the signals, as indicated on the horizontal axes. Both images credit: LIGO Scientific Collaboration.}
\label{fig:GW150914}
\end{figure}

It is time now to explain more completely our story's  final ingredient: Einstein's gravitational waves. The LIGO Scientific Collaboration, an international collaboration that has two detectors in the USA, made the first direct detection of gravitational waves on 14 September 2015 (Abbott et al 2016). The waves had been emitted a billion years ago by two very distant black holes. The holes, each around 30 times the mass of our Sun, had found themselves in a tight orbit around one another, maybe as a result of supernova explosions of both stars in a binary system. Their orbit  was in fact a doomed death spiral: their motion emitted gravitational waves, which carried energy away, which slowly but surely brought the holes closer and closer to each other. In fact, because of gravity's negative specific heat, the more energy the orbit lost, the faster the black holes moved in their orbits, and the stronger were the waves they radiated, which accelerated them to spiral in even faster. Such a dance can only end in catastrophe, and for these holes that happened when, after possibly several billions of years in orbit, they reached their last few orbits at exactly the right moment for their final radiated waves to travel a billion light-years and pass through the freshly-upgraded LIGO detectors when they were  in observation mode. LIGO's event, called GW150914, only told us about the final 0.2 seconds of this drama, as the holes made their last couple of orbits, merged to form a single much larger black hole, and then settled down into an eternally quiet state (left panel of \Figref{fig:GW150914}). The next big thing to happen to this hole is likely to be its final evaporation through the emission of  Hawking radiation, some $5\times 10^{72}$ years from now.

LIGO and its partner, the European Virgo detector near Pisa, have so far detected almost three hundred further black-hole mergers. But these events are not the ones that have played a role in the evolution of life. That distinction belongs to neutron-star mergers like GW170817 (right panel of \Figref{fig:GW150914}). Long ago, the supernovas of both stars in a binary system had led to two neutron stars in a tight orbit, and they reached the merger stage a mere 120 million years ago. Because their masses were much smaller than those of the black holes in GW150914,  their orbit evolved  more slowly, and the gravitational wave signal entered the observational frequency band of  LIGO and Virgo a couple of minutes before they merged, not fractions of a second. The merger produced a huge kilonova explosion that was observed by astronomers around the world, and across the electromagnetic spectrum: gamma rays, X-rays, visible light, infrared light,  radio waves. Most of the neutron-star material was not ejected, but rather merged into a single very massive neutron star and almost certainly collapsed into a black hole within a second or two. But the fraction that did get caught up in the explosion was blown away. What the astronomers' detailed observations have told us is that this exploding cloud of neutron-star material went through a sequence of nuclear processes that transformed it into an expanding cloud of the heaviest elements of the periodic table. 

Our earlier discussion of nuclear stability is useful now in helping us to understand why this was so. The material  blown off from the two stars started out as pretty much pure neutrons. Once free of the gravity that had stabilized them inside the neutron star, the neutrons  began to beta-decay back into protons and electrons, and the  protons' mutual repulsion began to break the material up, creating fission on a grand scale. The fragments were basically enormous neutron-rich clumps. But we know that no such \textquotedblleft clump\textquotedblright\ can be stable if it has more particles than the biggest (nearly-)stable nucleus, ${}^{238}$U. So these clumps kept breaking down hierarchically into smaller ones, a process that continued until the clumps were actually the nuclei we know from the periodic table. That is when the fragmentation stopped, because the nuclei were stable. It would not have gone to lower nuclei than ${}^{56}$Fe, because this is the most stable of all nuclei. 

Each fission event released a significant amount of energy in light particles, like neutrinos, photons, electrons, individual protons and electrons. And all of this energy from the nucleosynthesis was released in a mere fraction of a second. This energy made the explosion cloud start to expand at very high velocities, and after a few hours it was big enough and bright enough for astronomers to be able to see it. The spectra they recorded were consistent with the expectation that the cloud contained these heavy elements (Abbott 2017). It was our first glimpse of the synthesis of the upper half of the periodic table.

\begin{figure}
\includegraphics[height=3in]{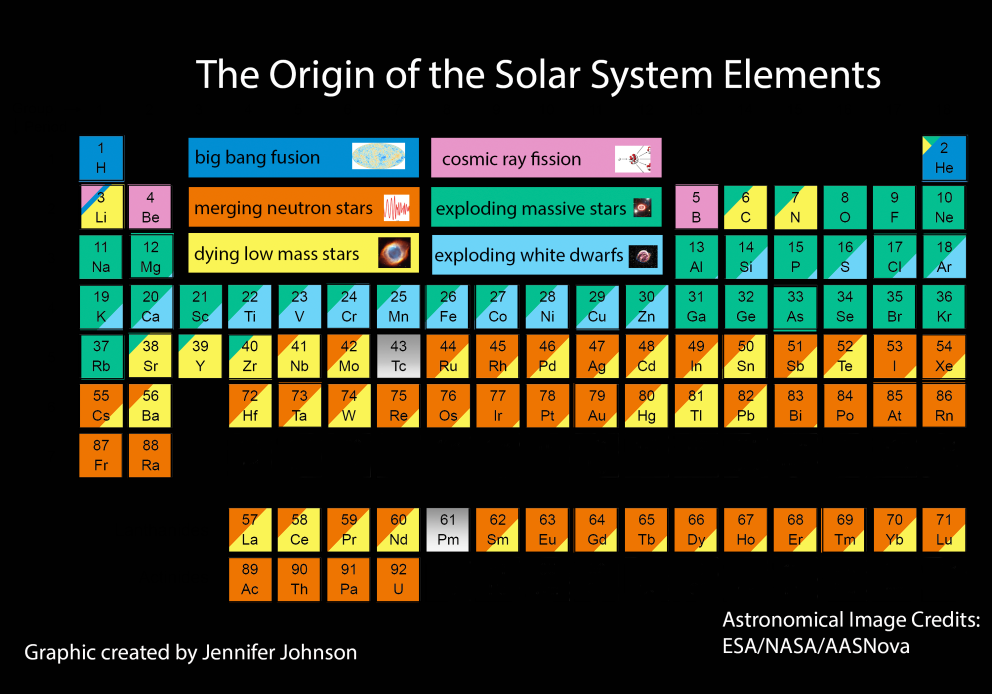}  
\caption{The principal sources of the elements. Credit: J Johnson, ESA/NASA/AASNova.}
\label{fig:nucleosynthesis}
\end{figure}

A kilonova like this is a rare event: it may occur in the Milky Way only once every 100,000 years or so. Supernovas, by contrast, happen at least once a century. The expanding kilonova cloud enriches the gas clouds significantly only in its near neighborhood in the Galaxy. The heavy elements in the primordial gas cloud from which the Sun eventually collapsed were therefore contributed over a long period of time by many events with much longer spacing in time. The remaining abundances of some of the relatively short-lived radioactive isotopes among these elements tell us that the last time our primordial cloud was ``seeded'' was about 100 million years before the Solar System formed from it. By contrast, its last enrichment by a supernova may have been  about 30 million years before the Sun began to form (O'Neill 2020). In fact, even shorter-lived isotopes, whose initial supply in the  Solar System would by now have completely decayed, continue to arrive at Earth in small quantities from  distant and more recent kilonovas, and this can be used to confirm the kilonova rate in the Galaxy that we quoted above (Hotekazaka, et al 2015). But our primordial cloud's abundances of all the stable elements in the periodic table, plus those of very long-lived radioactive isotopes like  the ones of interest here (${}^{238}$U  and ${}^{232}$Th), had been growing for many billions of years before the Solar System formed, as each nearby supernova or kilonova added  more to the cloud. See \Figref{fig:nucleosynthesis}.

Of course, it is important to ask whether there may also be other places where the heavier elements are synthesized. \Figref{fig:nucleosynthesis} gives an overview of how this works out for all the elements. One possibility for the heavier elements is a binary merger between a neutron star and a black hole. If the hole is not too massive, then the star may get shredded by the black hole's strong gravity before it enters the hole, and a similar process to the one just described would likely happen, and also requires gravitational waves. Other sources that do not require gravitational waves have been suggested, but they appear to contribute only a small fraction of the amount produced in mergers (Hotekazaka, et al 2018).

\section{Animal evolution in an unstable environment}

OK, now we are ready to ask and answer the question, So What? What do we humans need these elements for? Well, of course, our bodies use iodine, and this has led to the suggestion that gravitational waves were indeed necessary for life itself (Ellis et al 2024). But it is hard to be sure that Nature would not have found a different way to do what we need iodine for, if there had been no iodine on Earth. Less crucial for life itself, but central to human culture, is our love silver and gold jewelry; however, these elements are surely not essential for life. Nuclear reactors use uranium, but they are also not essential for life. What is often ignored is how important thorium and uranium have been for the preservation of life on Earth and for the evolution of advanced cognition in life on land (Schutz 2018, Piran 2019, Ellis et al 2024).

\begin{figure}
\includegraphics[height=3in]{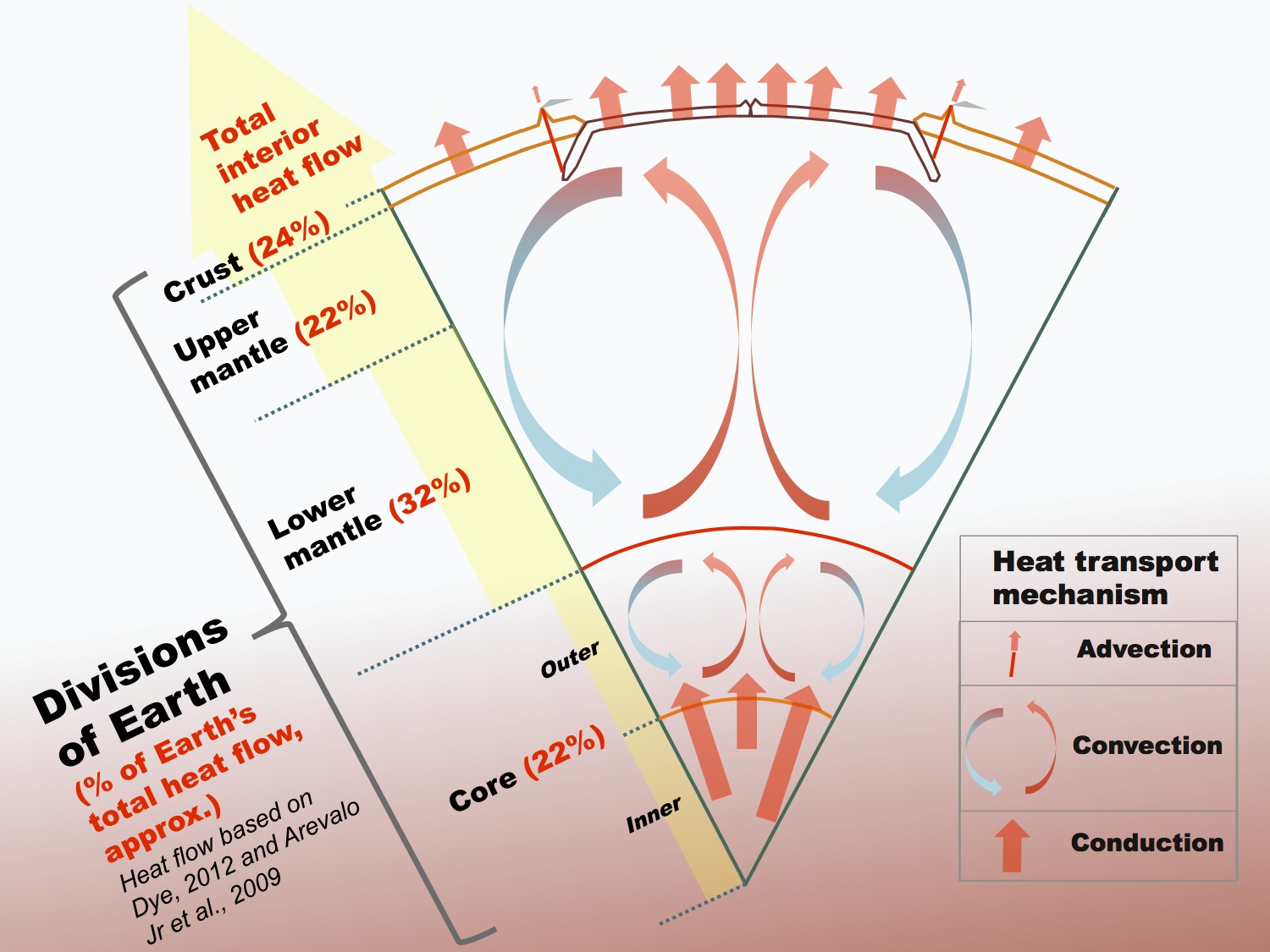}  
\caption{Schematic of Earth's heat flow. The main divisions of the Earth are shown, together with their relative contributions to the total heat flow to the surface, and the main transport mechanisms. Credit: Wikipedia, \textquotedblleft Earth's internal heat budget\textquotedblright.}
\label{fig:heatflux}
\end{figure}

They have influenced evolution through their importance in Earth's internal heat budget. At least half of the heat flowing out of Earth is generated by the decay of uranium and thorium (KamLAND Collaboration 2011, O'Neill 2020). Another important long-lived isotope is ${}^{40}$K, which was made by stellar processes that only required  Newtonian gravity. Its decays at present contribute perhaps another 10\% (O'Neill 2020). The fraction of the total heat flow that comes from radioactive decays is called the Urey Ratio, which for Earth  is therefore at least $0.6$. The rest of the heat flow is  primordial, just the slow cooling of our planet that has been going on since it was formed. What is more, it appears that the lower mantle generates significantly more heat than the upper mantle; this is the only reasonable explanation for the mantle convection that drives plate tectonics.(\Figref{fig:heatflux}). This extra heat must be radiogenic, since the primordial heat would have been more uniformly distributed. And geologists believe that this radiogenic contribution is vital, that if there were only the 40\% (or less) of heat that comes from cooling the  interior, then that would likely not be enough to drive plate tectonics. {\em The history of the movement of our continents, at least during the last 1 billion years, is due to the excess heat released deep in the mantle by} ${}^{238}$U {\em and} ${}^{232}$Th, {\em elements created by neutron stars brought into collision by their radiation of gravitational waves.} 

Tectonics is not the only effect that this heat flow has on our planet. The  temperature of Earth's core is high enough to keep iron molten, and Earth's rotation stirs up the liquid iron. This creates electric currents through the iron that generate Earth's magnetic field. If the core cools too much as the planet ages, then the iron could solidify, quenching the magnetic field. As we discuss below, if this had happened to Earth then it  would have had devastating consequences for life. And this may in fact have happened on Mars.  There is evidence on Mars for a primordial magnetic field that no longer exists (Acu\~na 1999). 

For the purpose of understanding the importance of gravitational-wave midwifed elements for life on Earth, the question is whether radioactive elements have helped to keep the iron core molten. The Urey Ratio, measured at Earth's surface, does not necessarily constrain the ratio of heat sources in the core, since geochemistry suggests that the uranium and thorium may be under-abundant there, while potassium might be present in larger abundances (O'Neill 2020). So at present there is not enough evidence to make the case one way or the other that radioactive uranium and thorium are keeping Earth's magnetic field strong. Since the question is open, we shall first briefly explain why the magnetic field is so important for life on Earth, after which we shall turn our attention to the issue of plate tectonics, where the importance of ${}^{238}$U and ${}^{232}$Th seems to be much clearer. 

{\bf Earth's magnetic field and the persistence of life.} For the last 4 billion years, Mars has had no large-scale planetary magnetic field (Acu\~na 1999). Mars is also a barren planet, with almost no atmosphere. These two facts are related. Earth's magnetic field shields our planet from the continuous stream of charged particles that we call the solar wind. The Sun's complex surface magnetic field, which is responsible for sunspots and solar prominences and solar storms, also constantly pushes away the outermost part of the solar atmosphere. Because of the high temperature of this gas, its particles are all ionized. We call this a plasma. When this plasma encounters Earth's magnetic field, the charged particles follow the field lines, so most of it is deflected by the field past the planet, while some is caught by the magnetic field lines and  channeled toward the magnetic poles, occasionally creating auroras at lower latitudes. 

If Earth had no magnetic field, the plasma would simply plow into its atmosphere,  over the course of billions of years gradually stripping  away not only the atmosphere but also the water in all the oceans. If, as seems possible, Mars was formed with an atmosphere similar to Earth's, then its initial magnetic field would have been a similar shield. But after the magnetic field died, the atmosphere and any surface water was blown away. If life had evolved in the watery regions of Mars in its early years, then once the magnetic field disappeared it would not have survived, at least not on Mars' surface. That is why the next searches for whatever may have passed for life on Mars will need to drill down deeper. The  discovery of a thriving biosphere full of microbial life at depths of a few hundred meters on Earth (Magnabosco 2018) have given added hope that something like this may still be present on Mars. 

It seems likely that, without the radioactive potassium, and also possibly the uranium and thorium whose synthesis was enabled by gravitational waves, Earth might well no longer have a magnetic field, an atmosphere, or liquid oceans. Absent the warming effect of the CO${}_2$ that our atmosphere has always carried, Earth's mean temperature would be below freezing. Terrestrial and oceanic life as we know it would, if it had had a chance to start at all, eventually have been annihilated, and it is possible that no form of life, even underground, would have survived. To make this association with gravitational waves clearer, however, better estimates of the amount of uranium and thorium in the deep core will be needed.

{\bf Plate tectonics and the evolution of higher cognition.} Given the uncertainty about what keeps Earth's magnetic field strong, it is possible that an Earth that had no uranium and thorium might still maintain a thriving community of life, but on continents that were no longer moving around. What might evolution have looked like on such a geologically quiet planet? To answer that hypothetical question, we first make a very brief sketch of what is known about  the 4 billion years of evolution on Earth as it really is, with those radioactive elements distributed through its body.

Earth formed about 4.6 billion years ago, and the first living cells seem to have appeared only 600 million years later. At that early time, Earth's interior was still very warm, both from its formation and then from the reheating caused by the impact at 4.5 billion years ago of a Mars-sized planet, the impact that led to the formation of the Moon from the debris (Young et al 2016). The radioactive isotopes would at that time have contributed only a small fraction of Earth's heat flow, so life would have formed whether the isotopes were there or not.

As Earth gradually cooled, continents formed and continental plates began moving. This is thought to have started sometime between 3 and 1 billion years ago. In the interval between 4 and 1 billion years ago, evolution created photosynthesis,  eukaryotic cells (cells with a membrane-bound nucleus), and mitochondria (which seem to have been essential for the later development of multicellular organisms). Earth's atmosphere began to become oxygenated as a result of photosynthesis around 2.5 billion years ago. By 1 billion years ago multi-cellular life had appeared. Importantly, by that time  the isotopes of uranium and thorium were already contributing the largest part of  Earth's outward heat flow, and  had become essential for maintaining the motions of the continents and for continued volcanism (O'Neill 2020). {\em The whole subsequent evolution of complex plants and animals occurred in an era in which geologic instability was being driven by the heat released by the decay of isotopes whose synthesis had been brought on by gravitational radiation.}

However, plate tectonics and volcanism have a much greater effect on terrestrial life and life in the shallows of continental margins, than they do on life in the body of the oceans. While the continents move (very slowly), the oceans remain one connected system that allows living things to relocate in response to this motion, to remain as much as possible in the environmental niche in which they evolved. It is therefore reasonable to think that, if gravity had been purely Newtonian, Earth would still have evolved its full complement of non-mammalian sea-life. But, possibly apart from the octopus, none of this life has  cognition approaching that of mammals. The most advanced ocean-dwelling animals are the cetacean mammals whose ancestors  adapted to the sea after evolving on land. 

The oceans provided the nursery for all animal life until the amphibians appeared 360 million years ago. The land animals that evolved after that did not experience a more nourishing or more supportive environment than fish have had. On the contrary, the air they lived in exposed them to the full impact of the environmental changes created by volcanoes, earthquakes, meteorite and asteroid impacts, mountain-building, continental collisions, and periods of extreme climate change. Challenges such as these force organisms to keep evolving and often drive them into new behaviors that are more sophisticated and more adaptive. 

It therefore seems reasonable to ask whether the challenges of an unstable geology were the main reason that cognition developed further on land than in the seas. Other occasional events, such as the famous impact of the asteroid that brought about the extinction of dinosaurs, should certainly be considered, but as we shall see the evidence is that their influences on the development of cognition were unimportant compared to that of tectonics. 

We shall see that, to understand human cognitive evolution, we need to consider one further recurrent cause of changes in climate: the cycle of ice ages. For the past 2.6 million years, these have come and gone with the regular changes in Earth's orbit that arise from Earth's gravitational interactions with the Moon and other planets. These orbital variations have  of course been present for Earth's entire 4.6 billion year history, so by themselves they don't explain these relatively recent ice ages. What has changed in the last few million years is that Earth is globally significantly cooler than earlier, so that the small changes in insolation that the orbital variations cause have been able just to tip the planet into and out of periods of substantial ice cover. Now, Earth's overall global temperature during the last few million years has been affected mostly by the tectonic motions of the plates. These change global oceanic circulation patterns, which redistribute the Sun's heat, and they move land masses toward or away from the poles, making it easier or harder to form ice sheets. Tectonics has put Earth into its current marginal temperature range, where  ice ages can happen. 


Having surveyed the main epochs of evolution, we turn now to the specific impact of these evolutionary pressures on cognition. The  group of animals with the most advanced cognition are the mammals. It is known, in particular, that mammalian brains have a special feature. Experiments show that even simple mammals, like rats, are able to make decisions by considering alternatives, based on their prior experiences. They appear to be able to imagine the consequences of pursuing each alternative available to them and, after a pause, to choose the one that is likely to provide the best outcome. In other words, they are able to consider before they act (Bennett 2023). Amphibians and reptiles, by contrast, also learn from experience but basically use this learning to guide immediate instinctive reactions to changing situations. 

It is not known when or why mammals developed this widely shared capability, but by itself it would not qualify mammals to be exceptionally intelligent. The size of the brain matters: the brain has to be able to remember enough relevant past experiences in order to use them to make good decisions as and when similar situations arise. And it turns out that brain size in mammals appears to have been more influenced by tectonic activity than by isolated events like the asteroid.

Mammals have been around since about 180 million years ago, but they got their chance to take over terrestrial habitats when the Chicxulub asteroid wiped out the dinosaurs 66 million years ago. At that time they were not particularly advanced cognitively, if we judge cognition by the ratio of brain size to body size (an admittedly crude criterion). Brains are energetically expensive organs to run, so an animal that puts only a small fraction of its body's energy budget into the brain is generally not going to get much out of it. At 100 million years ago, the typical mammalian brain-to-body size ratios were quite small, but they had started to increase slowly around 80 million years ago, long before the asteroid arrived. 

\begin{figure}
\includegraphics[height=3in]{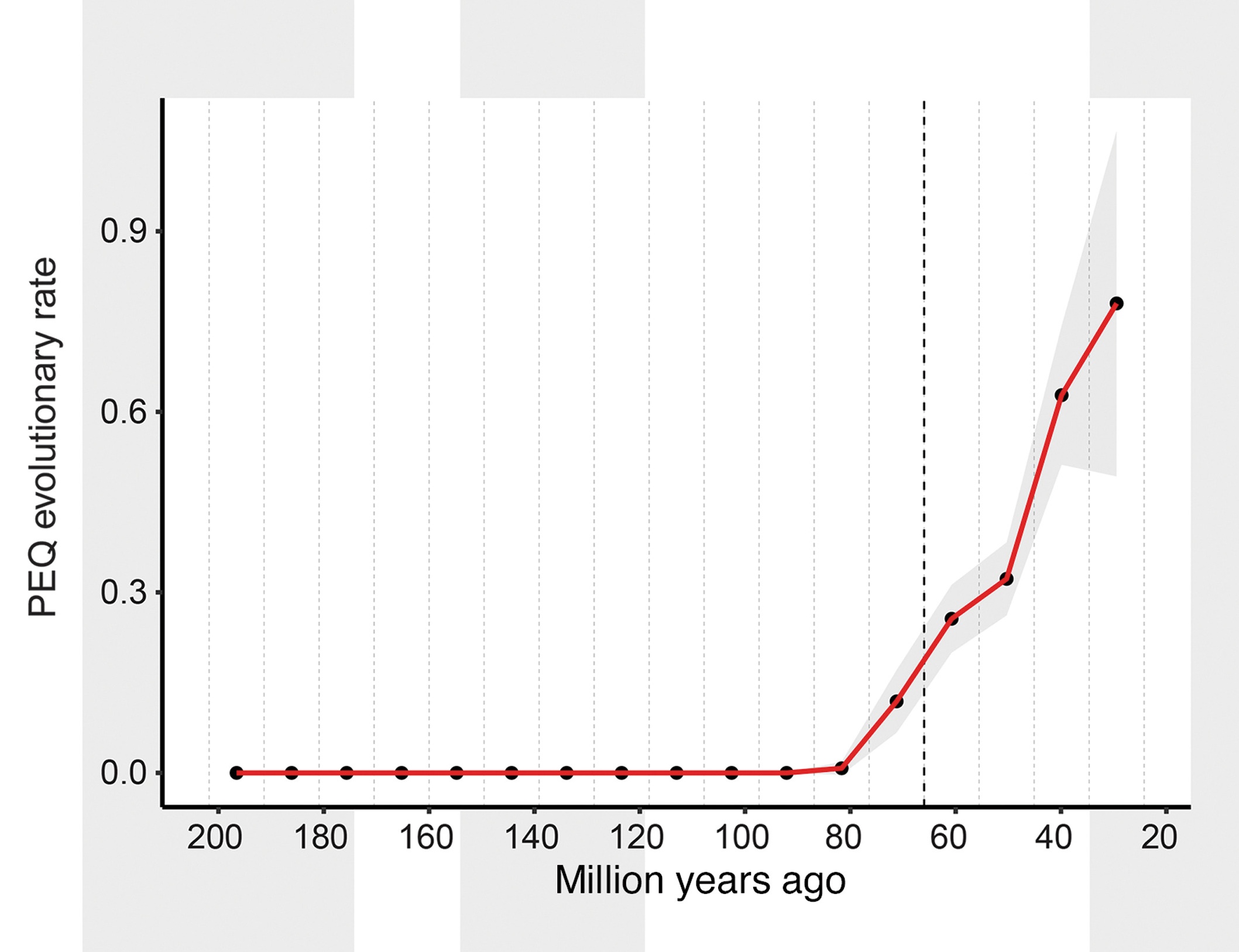}  
\caption{Rate of increase, per 10 million years, in the phylogenic encephalization quotient (see text for definition) of all mammals, plotted against time before present (in millions of years). The vertical dashed line marks the epoch of the asteroid impact.  Credit: Bertrand et al (2022).}
\label{fig:PEQ}
\end{figure}

What happened next is most clearly shown, not by looking at the simple and crude brain-to-body size ratio, but at the so-called PEQ, or phylogenetic encephalization quotient, which is the quotient of the actual brain-to-body size ratio to that expected from a fit to the distribution of this ratio among closely related species, normalized to a particular epoch (Bertrand et al 2022). This is believed to be much more closely related to relative cognitive improvements than the simple size ratio. The data gathered by Bertrand, et al (2022), for the average of a very large sampling of mammals, are shown in \Figref{fig:PEQ}. This plots the {\em  rate of change} of the PEQ vertically [in units of (10 million years)${}^{-1}$] against the look-back time from the present, plotted horizontally (in units of millions of years).

After about 80 million years ago, average mammalian PEQ continued a long period of slow increase, at a gradually increasing  rate. This slow acceleration in brain size continued steadily right through the asteroid event at 66 million years ago, showing no particular change at that time. What apparently did change rapidly after the asteroid was mammalian body size (not shown in this plot), as mammals quickly filled the empty habitats once dominated by dinosaurs. But the brain size apparently simply kept up with the body size. The African climate was much warmer than today's at that time and was very stable over tens of millions of years. Presumably in these favorable conditions, adapting to empty niches did not need rapid changes in cognition. 

But then, at 40  million years ago, the average mammalian PEQ rate of change increased sharply, and went even higher by 30 million years ago. This was toward the end of the Eocene, when global temperatures took a sharp drop. In Africa, open savannas replaced dense tropical forests. These changes in the African climate were caused by complex changes in the oceanic circulation, which in turn seems to have been initiated by a major tectonic event: the slow separation of Antarctica from Australia was completed around 30 million years ago. Before separation, the changing shape of this land mass would already have changed oceanic circulation, but full separation allowed water to flow between them, creating a rather rapid change in circulation and global climate. And after that, the icing up of Antarctica continued the cooling trend.  It would appear from \Figref{fig:PEQ} that this dramatic change in climate forced mammals in Africa to adapt in a major way to the changed habitat, and the ones who survived tended to have significantly bigger brains. This correlation between the inferred increase in mammalian cognition and tectonic activity is striking, even more so given the absence of a correlation of cognition changes with the impact of the asteroid. 

The evolution of the special cognition of humans also seems to have been driven by tectonic changes. Human cognition is built on that of primates, whose cognition had already increased more rapidly than most other mammals as they  adapted to  the stresses we have described. Humans separated from their last common primate ancestor around 7 million years ago. For the next 4 million years or so, our hominin ancestors remained rather small creatures with chimpanzee-sized brains, living on the margins of forests. But between 3 and 2 million years ago, the more primitive Australopithecus hominins gave way to those of the new genus Homo. The earliest of these, Homo habilis, had a transitional form, with a slightly larger brain and a flatter face. This is also when the first systematic manufacturing of stone tools, the Oldowan tool culture, began. By 2 million years ago, with the evolution of Homo erectus, our ancestors began to look more familiar: the brain size had almost doubled, the individuals were walking fully erect, and the tool culture (called Acheulean) had become more sophisticated, needing considerable practice to achieve proficiency in making the tools. If you met a group of Homo erectus individuals walking down the street, you would recognize them as human. In fact, erectus became the longest-lived human species of all time, lasting over a million years and spreading right across Africa and Eurasia. 

It is believed by specialists (e.g.\ Stanley 1998) that these evolutionary changes were driven, as before, by environmental changes. A long period of further cooling and drying had set in between 3 and 2 million years, in which savanna replaced much of the forest that Australopithecus had used for shelter. This change was accompanied by another: the climate became more unstable than before, with warmer wetter periods alternating with cooler dry periods, as indicated for example by lakes that come and go repeatedly in the geological record. These environmental stresses likely drove the rapid and steady increase in brain size, hence in cognition, that has accompanied human evolution from the time of H habilis, through H erectus, and to the eventual appearance of Homo sapiens about 300,000 years ago. 

And what caused these more recent environmental changes? The joining of North and South America culminated around 3 million years ago with the raising of the Isthmus of Panama, blocking oceanic currents that had previously flowed between the two continents. This resulted in part in a new circulation pattern east of Africa that was responsible for the major cooling of the continent. This pattern, still in place today, also brings regular monsoons to India, and drives the El Ni\~no Southern Oscillation. The cooling also tipped the temperature balance in the Northern Hemisphere, so that the periodic changes in Earth's orbit due to the Moon and planets now began to create alternating glacial and glacier-free epochs. It was these alternations which in turn caused the increased instability of Africa's climate. 

It is difficult to escape the conclusion that these tectonic events -- the separation of Australia from Antarctica and then the joining of North and South America -- were responsible for major environmental stresses in Africa that initiated the evolution, first of mammalian cognition to the general level at which we see it today, and then of human cognition. The details of these evolutionary processes were both of course very complex, but it seems unlikely that they would have happened at all if the motions of continents had not produced major changes in Earth's climate. This cognition is therefore another consequence of the heat released by the unstable isotopes of uranium and thorium, whose synthesis required that the physics of gravity include gravitational radiation.

\section{Perspective}

The production of ${}^{238}$U and ${}^{232}$Th in multiple kilonova explosions at the end of the gravitational-wave driven inspiral of two neutron stars or a neutron star and a black hole eventually had surprising consequences for life on Earth. Once incorporated in the body of the young planet, the isotopes provided a steady supply of heat from their radioactive decay. This heat is now helping to keep Earth's iron core molten, helping to sustain the protection that Earth's magnetic field gives from the solar wind. 

The heat from these isotopes today accounts for more than 50\% of Earth's heat flow, and this appears to be essential for sustaining plate tectonics and the other geological processes driven by mantle convection during the epoch of terrestrial life on Earth, and this has pushed mammals and specifically humans to the highest levels of cognition in the animal kingdom. Other environmental stresses, such as the comet impact 66 million years ago, have had major effects on evolution, but apparently not particularly on cognition. If Earth had cooled enough so that, even by as recently as 50 million years ago, the continents had frozen into some fixed configuration, then terrestrial life could probably still have developed a higher level of cognition than one finds in fish, but it would not have reached that of the average mammal today, to say nothing of the level of cognition that evolved in humans. 

This is an exceptional instance of the fine-tuning of the laws of physics, because without gravitational waves there would be life and probably even some form of mammalian life, but no humans cognitively advanced enough to understand the fine-tuning. Without Einstein's gravitational waves, we would not have had Einstein.

{\bf Acknowledgements.} BFS is grateful to the John Templeton Foundation for their hospitality at the discussion meeting, where an initial form of these ideas was presented and discussed. The approach in this paper has also benefited from discussions  with Kenta Hotokezaka, Avi Loeb, Ehud Nakar, Priya Natarajan, and Michael Paul.

\section*{References}

Abbott, B.~P.\, et al (2016). ``Observation of gravitational waves from a binary black hole merger'.' {\em Physical Review Letters}, {\bf 116}, 061102. 

Abbott, B.~P.\, et al (2017). ``Multi-messenger observations of a binary neutron star merger''. {\em Astrophysical Journal Letters}, {\bf 848}, L12.

Acu\~na, M.~H., et al (1999). ``Global Distribution of Crustal Magnetization Discovered by the Mars Global Surveyor MAG/ER Experiment''. {\em Science}
{\bf 284}, 790-793.

Bennett, M., (2023). {\em A brief history of intelligence} (William Collins).

Bertrand, O.~C. et al. (2022). ``Brawn before Brains in Placental Mammals after the End-Cretaceous Extinction''. {\em Science}, {\bf 376}, 80--85.

Carter, B. (1974). ``Large number coincidences and the anthropic principle in cosmology''. {\em IAU symposium 63: Confrontation of cosmological theories with observational data}, 291-298 (Dordrecht: Reidel). Republished online by Cambridge University Press, doi:10.1017/S0074180900235638 (7 Feb 2017). [For a survey of its complicated history, see ``Anthropic Principle'' in  Wikipedia\\  \verb|https://en.wikipedia.org/wiki/Anthropic_principle|\\ (visited 2025.02.13).]

Eichler, M., et al. 1989 {\em Nature} {\bf 340}, 126. 

Einstein, A. (1915). ``Feldgleichungen der Gravitation''. {\em Preussische Akademie der Wissenschaften, Sitzungsberichte},  (part 4), 844?847.

Einstein, A. (1918). ``Gravitationswellen''. {\em Preussische Akademie der Wissenschaften, Sitzungsberichte},  (part 1), 154?167.

Ellis, J., et al. (2024) ``Do we owe our existence to gravitational waves?''. {\it Physics Letters B}, {\bf 858}, 139028.

Gaia Collaboration (2018). ``Gaia Data Release 2: Mapping the Milky Way Disc Kinematics''. {\em Astronomy \& Astrophysics}, {\bf 616}, A11.

Hotokezaka, K., et al (2015) "Short-lived 244Pu points to compact binary mergers as sites for heavy r-process nucleosynthesis". {\em Nature Physics} {\bf 11} 1042.

Hotokezaka, K., et al (2018). ``Neutron star mergers as sites of r -process nucleosynthesis and short gamma-ray bursts''. {\em International Journal of Modern Physics D}, {\bf 27},  1842005.

KamLAND Collaboration (2011). ``Partial radiogenic heat model for Earth revealed by geoneutrino measurements''. {\em Nature Geoscience}, {\bf 4}, 647--651.

Lattimer, J.~M. and Schramm, D.~N.\ (1974). ``Black-hole-neutron-star collisions''. {\em Astrophysical Journal Letters}, {\bf 192}, L145--L147.

Magnabosco, C.,  et al (2018). ``The biomass and biodiversity of the continental subsurface''. {\em Nature Geoscience}, {\bf 11}, 707--717.

Metzger, B.~D.\ (2020). ``Kilonovae''. {\em Living Reviews in Relativity}, {\bf 23}, 1.

O'Neill, C., et al (2020). ``On the Distribution and Variation of Radioactive Heat
Producing Elements Within Meteorites, the Earth, and
Planets''. {\em Space Science Reviews}, {\bf 216}, 37.

Piran, T.\ (2019) "Extinction, $\Lambda$, GRBs, and GW", talk given at the meeting "From Deep Learning to the Dark Universe", London\\ \verb|https://www.ucl.ac.uk/astrophysics/seminars-and-events/deep-learning-dark-universe| (Visited 2025.03.01).

Schutz, B.~F.\ (2003). {\em Gravity from the ground up} (Cambridge University Press), Investigation 8.3, p.91.

Schutz, B.~F.\ (2018). ``The merger of two neutron stars, one year on: GW170817''. Blog post: \verb|https://bfschutz.com/2018/08/| (visited 2025.02.13).

Stanley, S.~M. (1998) {\em Children of the Ice Age} (Henry Holt and Co.).

Young, Edward D., et al (2016). ``Oxygen isotopic evidence for vigorous mixing during the Moon-forming giant impact''. {\em Science}, {\bf 351}, 6272.

\end{document}